%% file: IEEE-conference-template-062824.tex
\documentclass[conference]{IEEEtran}
\IEEEoverridecommandlockouts
\usepackage{cite}
\usepackage{amsmath,amssymb,amsfonts}
\usepackage{algorithmic}
\usepackage{graphicx}
\usepackage{xcolor}
\usepackage{biolinum}
\begin{document}

\title{Evaluating Chunking Strategies for Retrieval-Augmented Generation on Academic Texts\\

}

\author{\IEEEauthorblockN{Valentin J. J. Kreileder}
\IEEEauthorblockA{\textit{Computer Science} \\
\textit{Deggendorf Institute of Technology}\\
Deggendorf, Germany \\
kreileder@gmx.net}
\and
\IEEEauthorblockN{Johannes Reisinger}
\IEEEauthorblockA{\textit{Computer Science} \\
\textit{Deggendorf Institute of Technology}\\
Deggendorf, Germany \\
johannes.reisinger@th-deg.de}
\and
\IEEEauthorblockN{Andreas Fischer}
\IEEEauthorblockA{\textit{Computer Science} \\
\textit{Deggendorf Institute of Technology}\\
Deggendorf, Germany \\
andreas.fischer@th-deg.de}
}

\maketitle

\begin{abstract}
    Retrieval-Augmented Generation (RAG) systems use the question-answering capabilities of Large Language Models (LLMs) to access information outside their parameters.
    We evaluate if cluster-based semantic chunking improves retrieval and answer quality compared to fixed-size and recursive chunking evaluating on long,
    structured academic theses using the Retrieval Augmented Generation Assessment (RAGAs) framework.
    RAGAs based faithfulness shows limited reliability in this setup.
    Performance on fixed versus document specific questions varied substantially,
    likely related to the formatting of documents and preprocessing.
    Under the tested configuration, cluster-based chunking did not outperform simpler strategies.
\end{abstract}


\section{Introduction}\label{sec:introduction}
\input{sections/introduction}

\section{Related Work}\label{sec:related-work}
\input{sections/related-work}

\section{Methodology}\label{sec:methodology}
\input{sections/methodology}

\section{Results}\label{sec:results}
\input{sections/results}

\section{Conclusion \& Future Work} \label{sec:conclusion}
\input{sections/conclusion}

\section*{Acknowledgement}\label{sec:acknowledgement}
The first author used Claude (Opus 4.7) for a final grammar pass.
No text, figures, analysis, results, or conclusions were generated by the tool.
All technical content is the authors' own.

\bibliographystyle{IEEEtran}
\bibliography{IEEEabrv,references}

\end{document}

%% file: sections/introduction.tex
Large Language Models (LLMs) demonstrated impressive generative abilities,
yet their responses are limited by the information encoded in their parameters,
can suffer from hallucinations,
and do not show their inner workings.
Retrieval-Augmented Generation (RAG) systems address these issues~\cite{gaoRetrievalAugmentedGenerationLarge2024}.
Long documents cannot be processed as a whole because of the embedding models and LLMs context window,
therefore, those documents need to be split into smaller chunks.
These chunks are extracted from external documents, which can be done with different strategies.
Generated chunks are stored in a vector database before being retrieved with a user query.
The LLM uses the query and chunks to generate an answer.
The quality of the RAG-generated answer is coupled with the retrieval quality, the source data and the chunking strategy.
Conventional strategies are fixed-sized chunking or format based recursive chunking.
Semantic chunking gained prominence due to its potential to improve retrieval outcomes based on sentence similarity.
However, semantic strategies like cluster based chunking are computationally more demanding compared to traditional methods~\cite{quSemanticChunkingWorth2025}.
Core contributions of this paper are as follows:
\begin{itemize}
    \item We establish a measure combining faithfulness and answer relevancy called Answer Quality Score (AQS).
    \item We show issues regarding the evaluation using RAGAs on mid-range hardware.
    \item We show how different chunking strategies perform within this system using the RAGAs.
\end{itemize}

%% file: sections/related-work.tex
Qu et al. % Is Semantic chunking
demonstrate that plain fixed-size chunking is the most cost-effective approach~\cite{quSemanticChunkingWorth2025}.
Their experiments were conducted on a mixture of datasets including both original and artificially combined documents,
differing from the documents used in this work.
Their findings may not generalize to settings where documents are significantly larger and follow an internal structure.
We also use smaller sentence encoder under hardware constraints, further limiting comparison.
Late Chunking makes use of a long-context model to embed a full document,
then applies chunking and mean pooling to produce chunk vectors with better surrounding context than pipelines embedding after chunking~\cite{guntherLateChunkingContextual2025}.
Within the group's prior work, Reisinger et al.
propose document-level knowledge graphs from the hierarchical structure of chunk embeddings to detect document versions and plagiarism,
mitigating input- and context-conflicting hallucinations without a generative judge as done in this work~\cite{reisingerSemanticDocumentGraphs2025}.

%% file: sections/methodology.tex
\subsection{Chunking Strategies}\label{subsec:chunking-strategies}
Chunking splits data into segments an LLM can process within its context window,
and can be categorized in three main approaches: fixed-sized, format based, semantic chunking~\cite{reisingerSemanticDocumentGraphs2025}.
\begin{figure*}[t]
  \centering
  \includegraphics[width=0.8\textwidth]{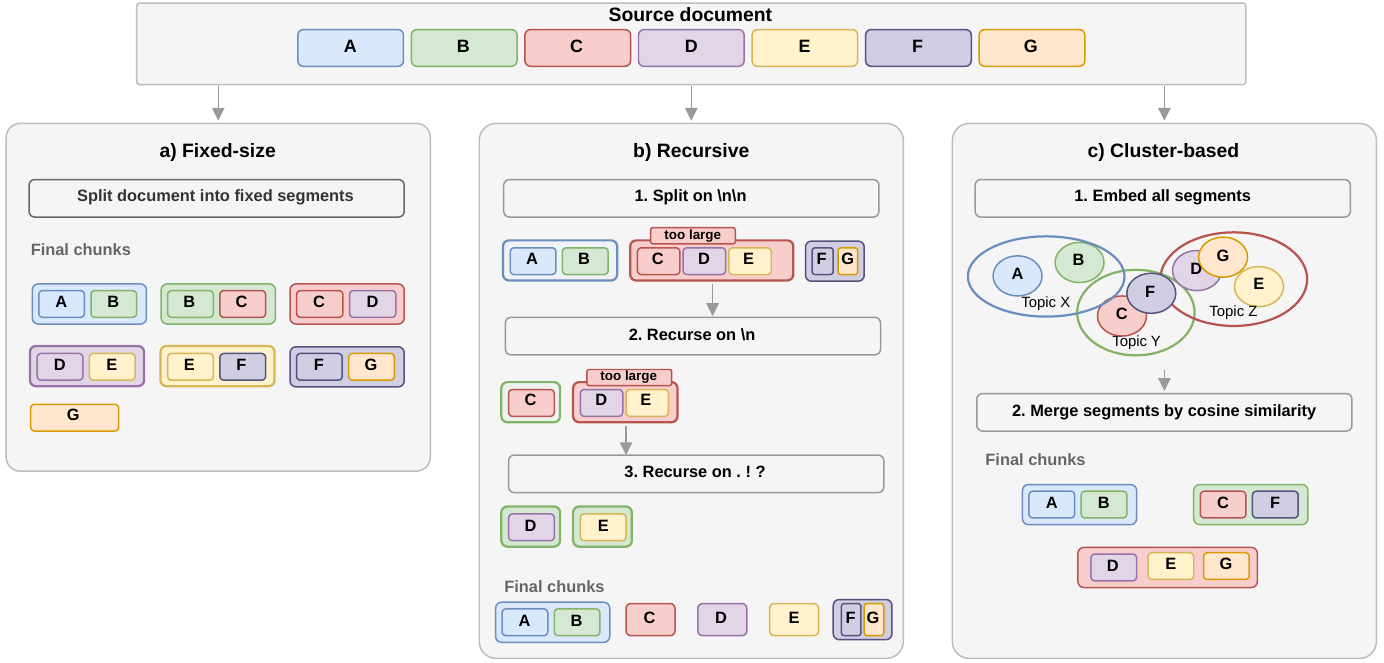}
  \caption{Used chunking methods.}
  \label{fig:chunking-categories}
\end{figure*}
\subsubsection*{Fixed-Sized Chunking}\label{subsubsec:fixed-sized-chunking}
Fig.~\ref{fig:chunking-categories} \emph{a)} depicts fixed-size chunking,
which splits the corpus into uniform-length segments.
Fixed boundaries ignore semantic structure,
so an overlapping sliding window is used in practice to reduce fragmentation across chunks~\cite{quSemanticChunkingWorth2025}.
The chunk sizes was 150 words with an overlap of 15.
\subsubsection*{Recursive Chunking}\label{subsubsec:recursive-chunking}
Fig.~\ref{fig:chunking-categories} \emph{b)} displays a format based strategy that splits the corpus into larger segments,
progressively subdividing those segments into smaller segments based on predefined criteria,
such as word limit or structural delimiters, combining the computational advantages of fixed-size chunking with format-aware flexibility.
Common delimiters include sentence boundaries,
or explicit markers such as newlines, paragraphs, or punctuation marks~\cite{reisingerSemanticDocumentGraphs2025}.
The selected chunk-size was 150 words.
\subsubsection*{Cluster-Based Chunking}\label{subsubsec:semantic-cluster-based-chunking}
Cluster-based chunking is a method where we combine semantically similar sentences to more coherent chunks (Fig.~\ref{fig:chunking-categories} \emph{c)}).
We follow~\cite{quSemanticChunkingWorth2025} but use \emph{all-MiniLM-L6-v2} instead of the larger encoders tested in~\cite{quSemanticChunkingWorth2025}, under hardware constraints.
We derive chunk size from target sentences and slack rather than a fixed cluster count,
and apply regex sentence splitting~\cite{SentencetransformersAllMiniLML6v2Hugging2024}.
The configuration is shown in Table~\ref{tab:chunker-sc-params}.
\begin{table}[h]
  \centering
  \caption{Clustering hyperparameters.}
  \label{tab:chunker-sc-params}
  \begin{tabular}{llr}
    \hline
    \textbf{Parameter} & \textbf{Description} & \textbf{Value} \\
    \hline
    $\lambda$            & positional/semantic weight & 0.25 \\
    $\tau$               & distance threshold         & 0.5  \\
    target\_sents        & sentences per chunk        & 3    \\
    slack                & max chunk size factor      & 1.2  \\
    method               & clustering algorithm       & single-linkage \\
    model                & sentence encoder           & all-MiniLM-L6-v2 \\
    \hline
  \end{tabular}
\end{table}
\subsection{RAGAs}\label{subsec:ragas}
For the experiment we use the Python framework RAGAs~\cite{esRAGAsAutomatedEvaluation2024}.
It makes use of an LLM to evaluate outputs based on prompt given by RAGAs,
reducing the need for human evaluation.
The evaluation pipeline is depicted in Fig.~\ref{fig:ragas-eval},
using the custom-made benchmark referred to in Section~\ref{subsec:dataset}.
Under VRAM constraints (16GiB), we selected llama3.2:3b as the generator, deepseek-r1:8b as the evaluator LLM, and all-MiniLM-L6-v2 as the embedder.
\subsection{Measures}\label{subsec:evaluation-measures}
\subsection*{TF-IDF bigram cosine similarity}
We use TF-IDF cosine similarity with bigrams all theses~\cite{manningIntroductionInformationRetrieval2008}.
Only bigrams that appear in at least two documents are taken into account.
The bigrams left are more domain-specific within the corpus.
Stop-words are removed before calculating the n-grams.
\subsection*{Context F1}
Context F1 is the harmonic mean between context recall and context precision~\cite{manningIntroductionInformationRetrieval2008}.
The RAGAs variants are used because they assess semantic relevance without requiring manually annotated ground-truth chunks~\cite{esRAGAsAutomatedEvaluation2024}.
\subsection*{Answer Quality Score}
We combine the scores faithfulness $F$ and answer relevancy $AR$ generated into one score named \emph{Answer Quality Score} (AQS).
Faithfulness measures the fraction of claims in the generated answer that are supported by the retrieved context.
RAGAs extracts and verifies them with an LLM.
Answer relevancy is the mean cosine similarity between the original question and a set of artificial questions generated by the LLM based on the response~\cite{esRAGAsAutomatedEvaluation2024}.
\\
The \emph{AQS}-score is defined as the harmonic mean between $F$ and $AR$ to penalize cases where either $F$ or $AR$ is low:

\[
  AQS=\frac{2 \cdot F \cdot AR}{F + AR}.
\]

This measure penalizes confident but unsupported answers with a high answer relevancy but low faithfulness.
\subsection{Dataset and QA set}\label{subsec:dataset}
\begin{figure}[ht]
  \centering
  \includegraphics[width=0.5\textwidth]{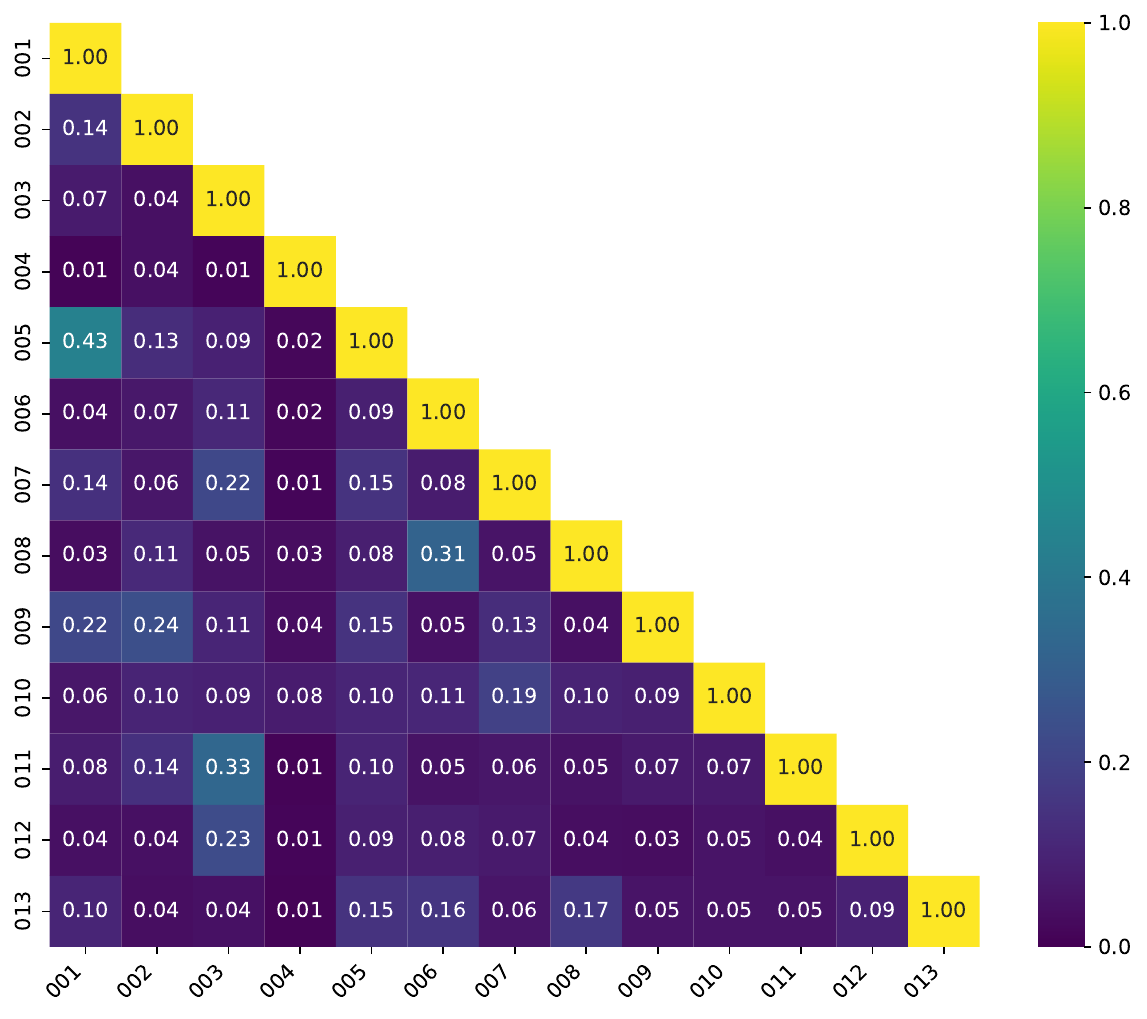}
  \caption{TF-IDF cosine similarity with bigrams as terms.}
  \label{fig:tf-idf-matrix}
\end{figure}

The dataset consists of thirteen theses, with a total of ten associated queries.
Five queries ask general questions about the author, title, and supervisors,
while the other five ask thesis-specific questions.
The theses mostly follow a faculty-standardized format.
The word-count ranges from 10{,}232 to 26{,}960 with a median of approximately 16{,}000.
No systematic relationship between document length and retrieval or answer performance was observed.
TF-IDF cosine similarity (as presented in Fig.~\ref{fig:tf-idf-matrix}) remains low to moderate across the corpus, indicating shared domain without substantial content overlap.
Outliers are document \emph{004}, where all comparative values are low, and documents \emph{001} and \emph{005}, which show the highest bigram similarity within the corpus.

%% file: sections/results.tex
\begin{figure}[ht]
  \centering
  \includegraphics[width=0.35\textwidth]{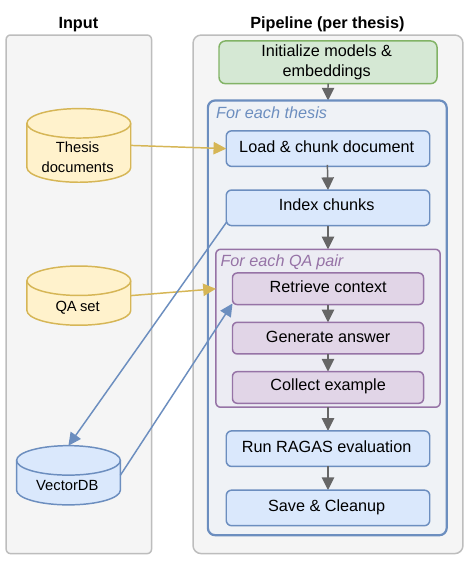}
  \caption{Pipeline concerning the evaluation of chunking methods using RAGAS}
  \label{fig:ragas-eval}
\end{figure}

We ran one evaluation using the configuration in Fig~\ref{fig:ragas-eval} producing 390 measurements.
Faithfulness calculation failed in 44\% of cases, % 47 recursive, 44 fs, 42 cluster based
compared to 2-3\% for the other metrics.
The evaluator either timed out or returned invalid values.
These empty values are spread relatively evenly across all chunkers,
with rates of 47\% for recursive chunking,
42\% for cluster-based chunking, and for ~44\% fixed-sized chunking.
We proceed with interpreting the results but with caution.
All measures are computed on valid samples,
resulting in a retained sample of around 97\%  for context F1 and  around 55\% for AQS.

\subsection*{Context F1 Results}
For \emph{fixed} questions, context f1 medians are 0 across all chunkers.
These first five questions target general information in the preliminaries.
Even after cleaning, preliminary artifacts and dot leaders survive,
polluting both indexing and retrieval.
For \emph{free} questions,
context f1 medians reach approximately 0.5 for recursive chunking and 0.3 for fixed-sized chunking outperforming the \emph{fixed} questions.
Cluster-based chunking still performs poorly,
though with a wider IQR  (interquartile range) not as close to zero as for the \emph{fixed} questions.
An example can be found in Fig.~\ref{fig:retrieval}.
\begin{figure}
\begin{quote}
  \footnotesize\ttfamily\raggedright
 \texttt 1.2 Research Objectives . . . . . . . . . . . . . . . . . . . . . . . . . . . . . . . . . 1 1.3 Research Questions . . . . . . . . . . . . . . . . . . . . . . . . . . . . . . . . . 1 1.4 Structure of the Thesis . . . . . . . . . . . . . . . . . . . . . . . . . . . . . . . 2 2 Background 3 2.1 Cybersecurity Landscape . . . . . . . . . . . . . . . . . . . . . . . . . . . . . . 3
\end{quote}\caption{Example: \emph{Fixed} question retrieval snippet}\label{fig:retrieval}
\end{figure}

\begin{figure*}[t]
  \centering
  \includegraphics[width=0.8\textwidth]{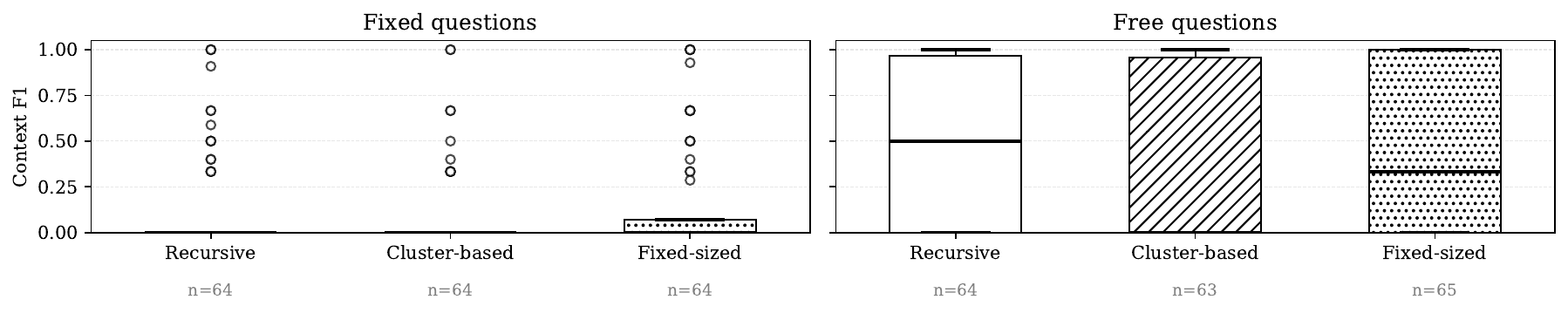}
  \caption{Boxplots displaying context F1 scores for the evaluated chunking strategies.}
  \label{fig:f1}
\end{figure*}

\subsection*{AQS Results}
\begin{figure*}[t]
  \centering
  \includegraphics[width=0.8\textwidth]{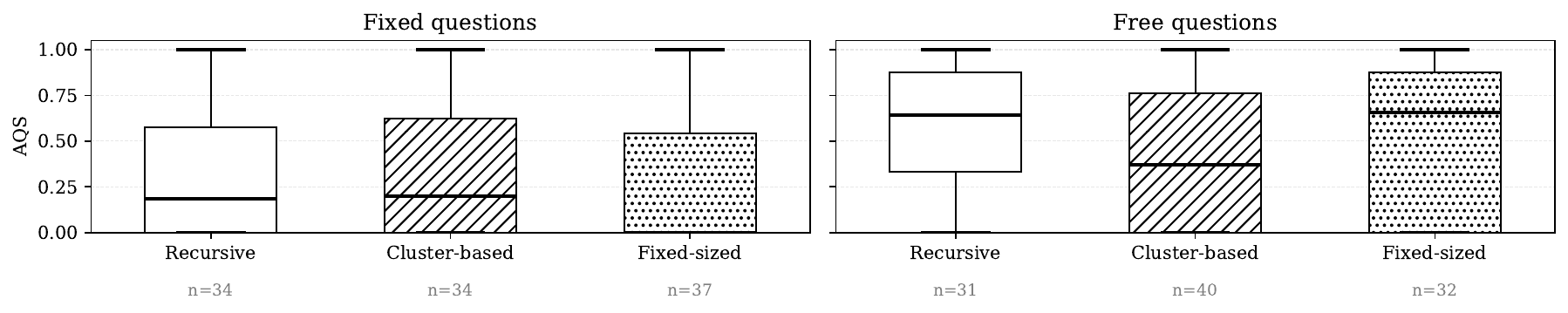}
  \caption{Boxplots displaying AQS scores for the evaluated chunking strategies.}
  \label{fig:fa}
\end{figure*}

AQS for \emph{fixed} questions in Fig.~\ref{fig:fa} stays low across all chunkers.
It is higher than the corresponding context f1 but still low,
with medians near zero and outliers reaching 1.00.
\emph{Free} questions outperform \emph{fixed} across all chunkers.
Fixed-sized and recursive chunking  both reach a median near 0.65,
with recursive showing the tightest IQR,
indicating more consistent generation quality.
Cluster-based chunking shows the lowest median of 0.40.

Retrieval and generation are decoupled in fixed questions where retrieval scores are near zero,
while generation scores remain moderate (as seen in Fig.~\ref{fig:f1},~\ref{fig:fa}),
suggesting that the judge finds some of the answers faithful and relevant,
despite the retrieved context being largely irrelevant.
This could be the case because the generator may answer using its parametric knowledge of generic thesis structure.
The RAGAs answer relevancy rewards question-answer independent of retrieved context, therefore resulting in high scores.
Also, narrowing reference contexts for fixed questions structurally punishes recall.

\subsection*{Discussion}\label{subsec:discussion}
This intends to reflect the capabilities of small-scale self-hosted RAG systems.
Using only academic theses,
while representative of long structured documents,
yields poor benchmark scores.

%% file: sections/conclusion.tex
This paper presented the performance of three chunking strategies on a corpus of academic theses using the RAGAs evaluation framework and mid-range hardware to limitations of a less cost-intensive environment,
and thus relied on small embedding, generation, and evaluation models.
The quality of retrieved chunks of \emph{fixed} questions as measured by RAGAs was limited, as seen in Fig.~\ref{fig:f1}, while being moderate for the \emph{free} questions.
The general answer quality was in parts good,
as seen in Fig.~\ref{fig:fa},
the results have to be interpreted with caution considering the loss of faithfulness.
This loss might be due to model capacity,
the highly specialized nature of the benchmark documents.
Across all configurations in this setup,
cluster-based semantic chunking did not yield any consistent improvement with the implemented configuration and adds computing complexity.
Simpler chunking strategies were overall more reliable though still did not yield reliable outputs,
with the best choice depending on the document provided.
\\
Future work could involve evaluating different clustering algorithms for cluster-based semantic chunking on the academic dataset and the use of bigger embedding models.
Instead of the RAGAs context recall proxy,
we will update the QA-set to be used with ID-based context recall.
Given the divergence between F1 and AQS on fixed questions,
a follow-up study should validate RAGAs measures against human-annotated chunk relevance to determine whether the gap reflects genuine retrieval issues or metric artifacts.